\documentclass{webofc}
\usepackage[varg]{txfonts}   % Web of Conferences font
\begin{document}
\title{Shift of Shapiro Step in High-Temperature Superconductor}

\author{\firstname{Kirill} \lastname{Kulikov}\inst{1,2}\fnsep\thanks{\email{kulikov@theor.jinr.ru}} \and
        \firstname{Yury} \lastname{Shukrinov}\inst{1,2} \and
        \firstname{Majed} \lastname{Nashaat}\inst{1,3} \and
        \firstname{Akinobu} \lastname{Irie}\inst{4}
        % etc.
}

\institute{BLTP, JINR, Dubna, Moscow Region, 141980, Russia 
\and
           State University of ``Dubna'', Dubna, Moscow region, 141980 Russia 
\and
           Department of Physics, Cairo University, Cairo, Egypt
\and
		   Department of Electrical and Electronic Systems Engineering, Utsunomiya University, Japan
          }

\abstract{%
Influence of the charge imbalance effect on the system of intrinsic Josephson junctions of high temperature superconductors under external electromagnetic radiation are investigated. We demonstrate that the charge imbalance is responsible for a slope in the Shapiro step in the IV-characteristic. The nonperiodic boundary conditions shift the Shapiro step from the canonical position which determined by   a frequency of external radiation. We also demonstrate how the system parameters affect on the shift of Shapiro step. 
}
\maketitle
\section{Introduction}
\label{intro}

The phase dynamics of the layered superconducting materials have attracted a great interest because of rich and interesting physics from one side and perspective of applications from the other one [1, 2]. In particular, the nonequilibrium effects created by stationary current injection in high-$T_c$ materials have been studied very intensively in recent years \cite{Artemenko97,Preis98,Shafranjuk99, Helm00,Helm01,Helm02,Bulaevskii02}. However,  the charge imbalance in the systematic perturbation theory is considered only indirectly as far as it is induced by fluctuations of the scalar potential \cite{Artemenko97, Preis98,Helm01}. In Ref.\cite{keller05}, it is taken into account as an independent degree of freedom and, therefore, the results are different from those of earlier treatments. In addition, due to the fact that charge does not screen in the superconducting layers a system form an intrinsic Josephson junctions (IJJ) \cite{Koyama96,Dmitry}. \textit{Such system cannot be in the equilibrium state at any value of the electrical current}.  The influence of charge coupling on Josephson plasma oscillations was stressed in Refs.\cite{Koyama96, Helm02}. 

Last few years, two theoretical models are widely used to describe IJJ: capacitively coupled Josephson junctions (CCJJ) model and charge imbalance (CIB) model. In CCJJ model a non-vanishing generalized scalar potential appears due to the breaking of charge neutrality, but in CIB model it is related to the quasiparticle charge imbalance as well. Actually, relaxation length of charge imbalance in the layered system could be much larger than any other characteristic lengths. Therefore, both effects could exist in HTSC simultaneously because the thickness of superconducting layers is smaller than the Debye length and thus obviously less than characteristic length of disequilibrium relaxation. In current paper, we study the nonequilibrium effects created by current injection in a stack of IJJ under external electromagnetic radiation.

\section{Model}
\begin{figure}[htb]
 \centering
\includegraphics[width=80mm]{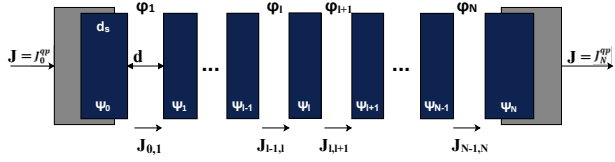}
\caption{Layered system of $N+1$ superconducting layers forms a stack of Josephson junctions. Since the 0-th and N-th layers are in contact with normal metal, their thicknesses $d^0_s$ and $d^N_s$  are different from the thickness of the other S-layers $d_s$ inside of the stack due to the proximity effect.}
 \label{fig:layer}
\end{figure}

A system of $N+1$ superconducting layers (S-layers) presented in Fig. \ref{fig:layer} is characterized by the order parameter $\Delta_l(t) = \vert \Delta \vert \exp(i\theta_l(t))$ and time-dependent phase $\theta_l(t)$.
The thickness of the S-layer is comparable with the Debay screening $r_D$ length that leads to the generalized Josephson relation \cite{epl}. The total current density $J_l$ through each S-layer is given as a sum of displacement, superconducting, quasiparticle, diffusion and nonequilibrium terms. Those equations together with kinetic equations for the nonequilibrium potential $\Psi_l(t)$ describe the physics of IJJs in HTSC. In the dimensionless form the system of equations are
\begin{eqnarray}
 	\dot{v}_{l}&=&  \bigg{[} I -\sin\varphi_{l}-\beta\dot{\varphi}_{l} + A \sin{\omega} \tau + I_{noise}+\psi_{l}-\psi_{l-1}\bigg{]} \\
 	\dot{\varphi}_{1} &=& v_{1} - \alpha (v_{2}-(1+\gamma)v_{1}) +\frac{\psi_{1}-\psi_{0}}{\beta} \\
 	\dot{\varphi}_{l} &=& (1+2\alpha)v_{l} - \alpha (v_{l-1}+v_{l+1}) +\frac{\psi_{l}-\psi_{l-1}}{\beta} \\
 	\dot{\varphi}_{N}&=& v_{N} - \alpha (v_{N-1}-(1+\gamma)v_{N}) +\frac{\psi_{N}-\psi_{N-1}}{\beta} \hspace{0.5cm} \\
 	\zeta_{0} \dot{\psi}_{0} &=& \eta_{0}  \left( I + A \sin{\omega} \tau - \beta \dot{\varphi}_{0,1} + \psi_{1}-\psi_{0}\right) -\psi_{0} \\
 	\zeta_{l} \dot{\psi}_{l}&=& \eta_{l} ( \beta[\dot{\varphi}_{l-1,l} - \dot{\varphi}_{l,l+1} ]+ \psi_{1-1}+\psi_{l+1} - 2\psi_{l})  - \psi_{l} \nonumber \\
 	\zeta_{N} \dot{\psi}_{N} &=& \eta_{N}  \left( -I - A \sin{\omega} \tau + \beta\dot{\varphi}_{N-1,N} + \psi_{N-1}-\psi_{N}\right) -\psi_{N} \hspace{0.3cm}
 	\label{eq:8}
 \end{eqnarray}
where the dot shows a derivative with respect to $\tau=\omega_pt$, $v_l$ between the layers $l-1$ and $l$, $ v_l(t) \equiv v_{l,l-1}(t) $, $\varphi_{l}(t)$ is phase difference across the layers $l-1$ and $l$, $\alpha = \epsilon\epsilon_{o} /2e^{2} N(0) d $ is the coupling parameter, $\epsilon$ is the dielectric constant, $\epsilon_{o}$ is the vacuum permittivity, $d$ is the distance between the superconducting layers and $N(0)$ is the density of states,  $I=J/J_{c}$ is the dimensionless current density, $J_{c}$ is the critical current density, $\omega_p=\sqrt{\frac{2eJ_c}{\hbar C}}$ is plasma frequency, $C$ is the capacitance. Other dimensionless parameters are the dissipation parameter $\beta=\frac{\hbar\omega_p}{2eRI_c}$, $R$ is the junction resistance, the normalized quasiparticle relaxation time $\zeta_{l}=\omega_p \tau_{qp}$ and the nonequilibrium parameter $\eta_{l}=\frac{4\pi r_D^2\tau_{qp}}{d^{l}_{s}R}$, where $d^{i}_{s}$ is the thickness of the S-layers, and $\tau_{qp}$ is the quasiparticle relaxation time. The parameter of the nonperiodic boundary conditions $\gamma$ is $\gamma=\frac{d_s}{d_s^0}=\frac{d_s}{d_s^N}$. The term $A \sin \omega \tau$ introduces the effect of external radiation with amplitude A and frequency $\omega$, which are normalized to $J_{c}$ and $\omega_{p}$, respectively. To reflect the experimental situation, we have added the noise $I_{noise}$ in the bias current with the amplitude $\sim 10^{-8}$ which is produced by random number generator and its amplitude is normalized to the critical current density value $J_{c}$.

This system of equations is solved numerically using the fourth order Runge-Kutta method. We assume that due to the proximity effect the thickness of the first and the last S-layer larger than the middle one. Therefore, the nonequilibrium parameters depend on the parameter of boundary conditions $\gamma$, $\eta_{0,N}=\gamma\eta_l $, where $l=1,2,..,N-1$. We consider the underdamped case with the McCumber parameter $\beta_{c}=25$ or $\beta=0.2$.

\section{Results}

In Ref.\cite{epl} we have shown that in the system of intrinsic Josephson junctions of high temperature superconductors under external electromagnetic radiation charge imbalance is responsible for a slope in the Shapiro step in the IV-characteristic. The value of the slope increases with a nonequilibrium parameter. We demonstrate that coupling between junctions leads to the distribution of the slope's values along the stack. It was shown also that the nonperiodic boundary conditions shift the Shapiro step from the canonical position.

\begin{figure}[h!]
	\centering
	\includegraphics[width=50mm]{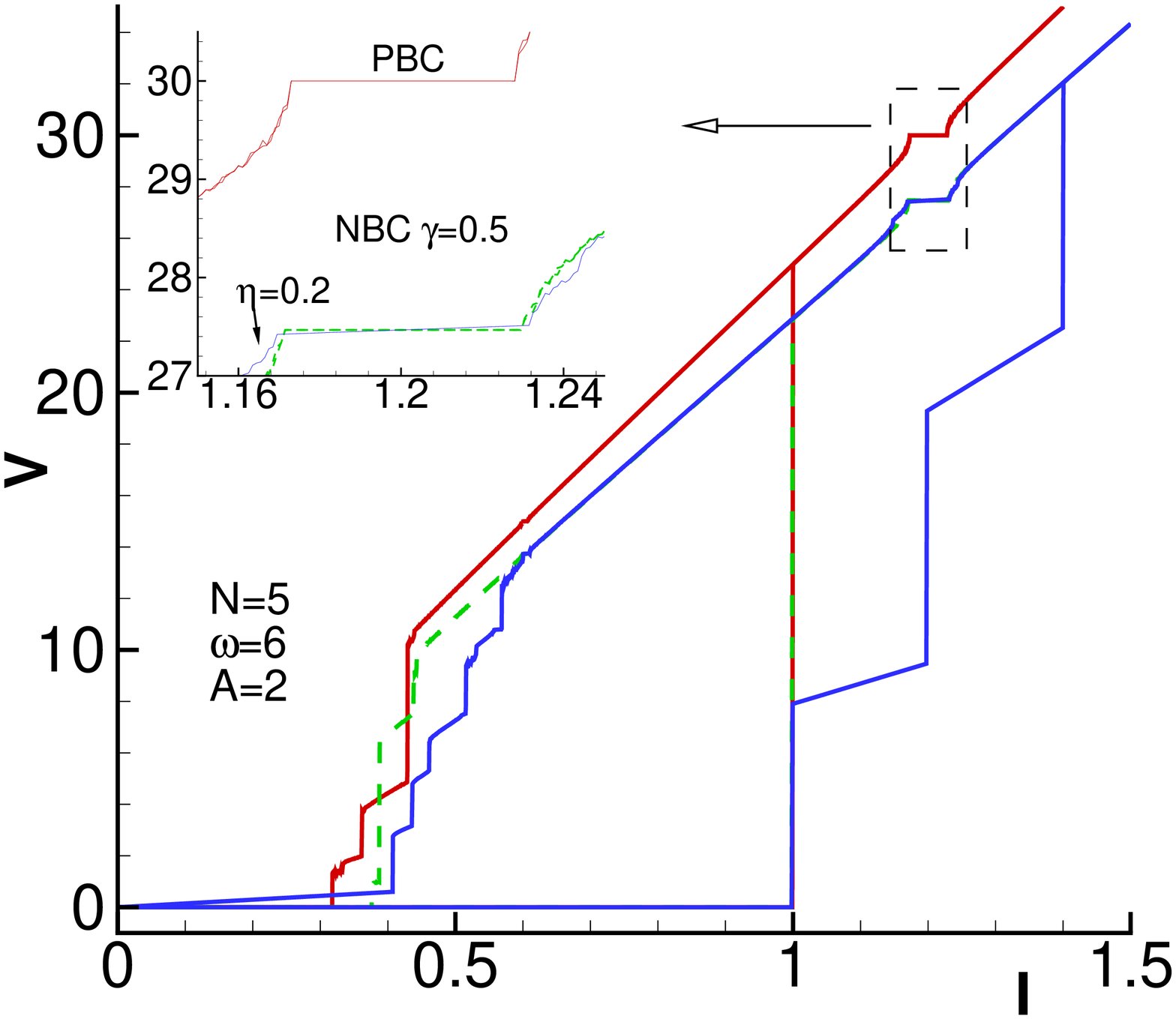}
	\caption {The IV-characteristics of JJ stacks without the charge imbalance $\eta=0$ (dashed line) and at $\eta=0.2$ (solid line). The  results for the periodic boundary condition are shown for comparison.  The enlarged parts of the IV-characteristics with the SS are shown in the inset.}
	\label{fig:2}
\end{figure}

The simulated IV-characteristics of JJs stack in the case without the charge imbalance $\eta=0$ (dashed line) and at $\eta=0.2$ (solid line) are presented in Fig.\ref{fig:2}.
The IV-curve without the charge imbalance at periodic boundary conditions are shown as well. The simulations have been made for the stacks with five JJs, coupling parameter $\alpha=0.5$ and $\gamma=0.5$.  We see that the position of the SS (at periodic boundary condition (PBC)) corresponds to the canonical value of the SS voltage  $V=30$ in agreement with the value of external frequency  $\omega=6$  and a number of junctions in the stack $N=5$. The nonperiodic boundary conditions with  $\gamma\neq0$ shift the outermost branch relatively to the curve at PBC, leading to the corresponding shift of the Shapiro steps. The charge imbalance manifests itself as appearance of the slope in the Shapiro step, which is clearly demonstrated in the inset for the case $\eta=0.2$.

\begin{figure}[htb]
 \centering
\includegraphics[width=50mm]{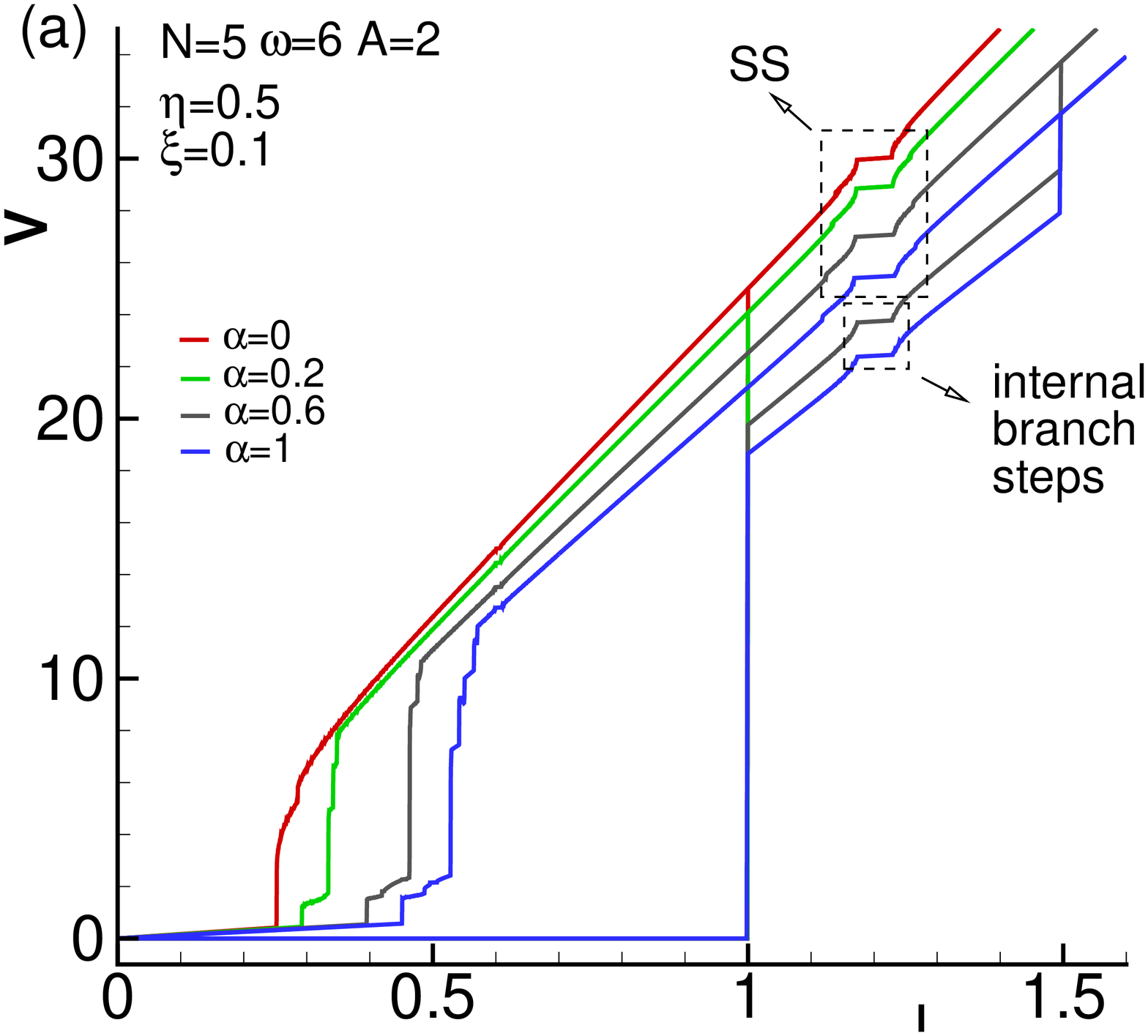}\includegraphics[width=50mm]{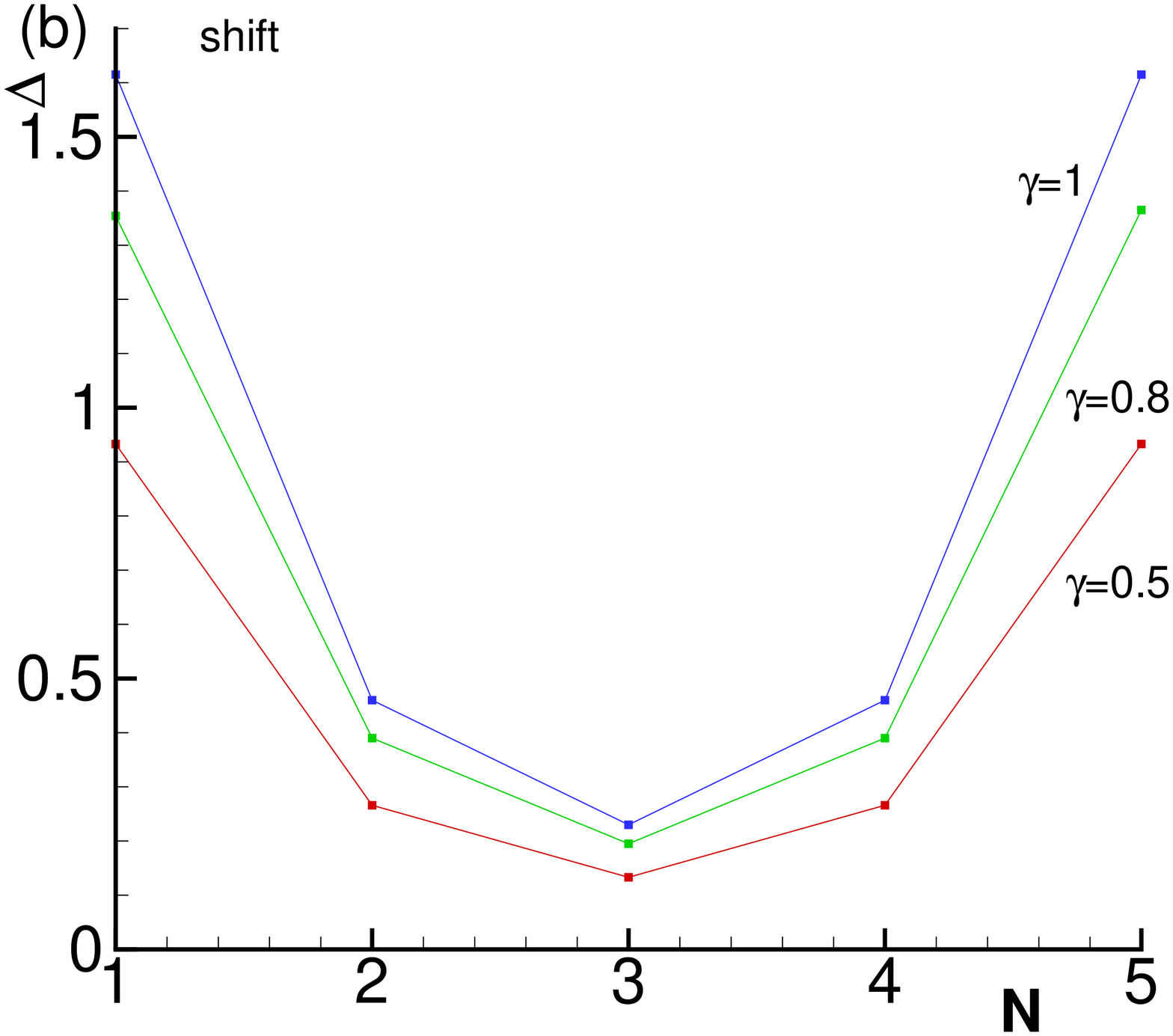}
\caption{(a) IV-characteristics of the stack of $N=5$ JJ at different coupling parameter $\alpha=0, 0.2, 0.6, 1$. (b) Distribution of the SS shift along the stack of $N=5$ JJ at different parameter of $\gamma=0.5, 0.8, 1$.}
 \label{fig2_dist_dif_gamma}
\end{figure}

Influence of the coupling parameter on the shift of Shapiro step is shown in Fig.~\ref{fig2_dist_dif_gamma}(a). Increasing of $\alpha$ leads to increasing of the shift value. The steps on the IV-characteristic at $\alpha=0, 0.2, 0.6, 1$ is indicated by the large dashed rectangle. The IV-characteristics at large $\alpha$ demonstrate also additional steps those appears on an internal branch. Those Shapiro steps indicated by the small dashed rectangle.

The shift of the Shapiro step depends on the value of $\gamma$ and on the coupling parameter $\alpha$. Fig.~\ref{fig2_dist_dif_gamma}(b) shows the distribution of shift of step along the stack with parameters $\gamma=0.5, 0.8, 1$. One can see that the maximum of the shift value represents first and last Josephson junction. Thickness of superconductive layers of those junctions are larger than the others. The distribution of the shift can be also seen in the middle layers due to the coupling between the Josephson junctions.

Thus, the Shapiro step demonstrates a shift of its position from the canonical value $N \omega$, where $N$ is the number of junctions in the stack and $\omega$ is the frequency of the external radiation. The value of this shift depends on the boundary conditions and coupling between Josephson junctions. Due to the coupling, the effect of the boundary conditions is extended to the neighboring junctions.

\section{Acknowledges}
The reported study was funded by RFBR according to the research projects 15--29--01217 and 16--52--45011\_ India.

%
% BibTeX or Biber users please use (the style is already called in the class, ensure that the "woc.bst" style is in your local directory)
% \bibliography{name or your bibliography database}

\begin{thebibliography}{}
\bibitem{Artemenko97} S. Artemenko and A. Kobelkov, Phys. Rev. Lett. {\bf 78},  3551  (1997).
	\bibitem{Preis98} C. Preis, C. Helm, J. Keller, A. Sergeev, and R. Kleiner, in {\it Superconducting Superlattices II: Native and Artificial}. I. Bozovic and D. Pavona, editors, Proceedings of SPIE Volume {\bf 3480},  236  (1998).
	\bibitem{Shafranjuk99} S. E. Shafranjuk, M. Tachiki, Phys. Rev. B {\bf 59}, 14087 (1999).
	\bibitem{Helm00} C. Helm, C. Preis, C. Walter, J. Keller, Phys. Rev. B {\bf 62}, 6002 (2000).
	\bibitem{Helm01} C. Helm, J. Keller, C. Preis, and A. Sergeev, Physica C {\bf 362}, 43 (2001).
	\bibitem{Helm02} C. Helm, L. N. Bulaevskii, E. M. Chudnovsky, M. P. Maley, Phys. Rev. Lett. {\bf 89}, 057003 (2002).
	\bibitem{Bulaevskii02} L. N. Bulaevskii, C. Helm, A. R. Bishop, M. P. Maley, Europhys Lett. {\bf 58}, 057003 (2002).
	\bibitem{keller05} J. Keller, D. A. Ryndyk, Phys. Rev. B {\bf 71}, 054507 (2005).
	\bibitem{Koyama96} T. Koyama and M. Tachiki, Phys. Rev. B {\bf 54}, 16183 (1996)
	\bibitem{Dmitry} D. A. Ryndyk,  Phys. Rev. Lett. {\bf 80}, 3376 (1998).
	\bibitem{Dmitry} D. A. Ryndyk,  Phys. Rev. Lett. {\bf 80}, 3376 (1998).
	\bibitem{epl} Yu. M. Shukrinov, M. Nashaat, K. V. Kulikov, R. Dawood, H. El Samman and Th. M. El Sherbini, EPL {\bf 115}, 20003 (2016)
\end{thebibliography}

\end{document}